\documentclass[ajl]{emulateapj}
\usepackage{natbib}
\usepackage{pifont}
\usepackage{color}
\usepackage{graphicx}
\usepackage{epsfig}

\usepackage[usenames,dvipsnames,svgnames,table]{xcolor}

\shorttitle{Growth of supra-arcades  fans}
\shortauthors{Innes et al.}

\newcommand{\degree}{\ensuremath{^\circ}}
\newcommand{\kms}{km~s$^{-1}$}

\newcommand{\FeXXI}{\ion{Fe}{21}}

\begin{document}
\doublespace

\title{Observations of supra-arcade fans: instabilities at the head of reconnection
jets}

\author{D. E. Innes \altaffilmark{1,2}}
\email{innes@mps.mpg.de}

\author{L.-J. Guo\altaffilmark{1,2}}

\author{A. Bhattacharjee\altaffilmark{2,3,4}}

\author{Y.-M. Huang\altaffilmark{2,3,4}}

\author{D. Schmit\altaffilmark{1,2}}

\affil{$^1$Max Planck Institute for Solar System Research, 37077
G\"ottingen, Germany}
\affil{$^2$Max Planck/Princeton Center for Plasma
Physics, Princeton, NJ 08540, USA}
\affil{$^3$Department of
Astrophysical Sciences and Princeton Plasma Physics Laboratory, Princeton
University, Princeton, NJ 08540, USA}
\affil{$^4$Center for Integrated Computation and Analysis of Reconnection and
Turbulence and Space Science Center, University of New Hampshire, Durham, NH 03824, USA}

\begin{abstract}
Supra-arcade  fans are bright, irregular regions of emission that
develop during eruptive flares, above flare arcades.
The underlying flare arcades are thought to be a
consequence of magnetic reconnection
along a current sheet in the corona.
At the same time, theory predicts plasma jets from the reconnection site
 which would be
extremely difficult to observe directly becuase of their low density. It has been suggested that
the  dark supra-arcade downflows (SADs) seen falling through  supra-arcade
fans may be low density jet plasma. The head of a low density
jet directed towards higher density plasma would be
Rayleigh-Taylor unstable, and lead to the development of
rapidly growing low and high density fingers along the interface.
Using SDO/AIA 131\AA\ images, we show details of SADs
seen from three different orientations with respect to the flare arcade and current
sheet, and highlight features that have been previously unexplained, such as the
splitting of SADs at their heads,  but are a natural consequence
of instabilities above the arcade.
Comparison with 3-D magnetohydrodynamic simulations suggests that
supra-arcade downflows are the result of secondary instabilities
of the Rayleigh-Taylor type in the exhaust of reconnection jets.
\end{abstract}


\section{Introduction}

Supra-arcade fans are diffuse regions of hot, 10~MK, plasma seen above flare
arcades \citep{Svestka98,McKenzie99,McKenzie00,Hanneman14}.
The supra-arcade fan can  extend up to about
150~Mm above the flare arcade and can last for several hours.
Often dark, tadpole-shaped structures, known as supra-arcade downflows
(SADs),
 are seen descending through the
bright fans, and the top of the arcade
appears spiky \citep{McKenzie00,Innes03a,Asai04,Verw05,Savage11}.
The SADs which are low density plasma \citep{Innes03a,Savage12}  are
believed to be a consequence of reconnection high in the corona because,
among other reasons,
they correlate well with bursts of hard X-rays \citep{Asai04,Khan07}.
Understanding the
cause of SADs could shed light on the reconnection process itself
as well as other possible secondary instabilities. One
difficulty is that SADs are only visible where there is
sufficient contrast with fan emission, so their structure outside
the fan is not known. Nevertheless high resolution observations of the interaction
between the bright fan plasma and the SADs can possibly discriminate between proposed
scenarios.

Supra-arcade fans generally extend along the whole length of the underlying arcade,
and are thought to be flat sheet-like structures or the narrow, low density
SADs would not show up as dark streaks.
When observed above cusp-shaped flare arcades, they have been interpreted as
 current sheets \citep{Savage10,LiuW13,Liu13}.
These current sheets, seen as rays of increased density and temperature \citep{Ko03, Ciar08},
 separate regions of oppositely
directed magnetic field above newly formed flare loops.
The
current sheets are expected to be seen edge-on above arcades seen from one end so that they
appear cusp-shaped, and
face-on above arcades seen from the side.

SADs are best seen when the current sheet is face-on
\citep{Savage10,Warren11}. A few have also been seen below
an almost edge-on current sheet \citep{Liu13,LiuW13}.
In high resolution extreme ultraviolet (EUV) images their appearance changes as the
supra-arcade fan develops (see animations in \citet{Warren11}).
Early downflows are often broader and shorter
than the more distinctive, tadpole-shaped flows that are
seen once the supra-arcade fan has
fully developed.


The SADs often appear above and  occasionally alongside rapidly retracting loops, known as
supra-arcade downflow loops (SADLs) with similar velocity (50-400~\kms),
acceleration and height
\citep{Savage11}.
SADLs appear bright against the dark corona
and SADs are darker than the surrounding supra-arcade fan plasma.
It has been noted that contrary to SADs, SADLs are more likely to be seen
when the flare arcade
is oriented with its axis parallel to the line-of-sight
\citep{Savage10, Savage11, Warren11,LiuW13,Liu13}.

Early analyses of SADs concluded that they were evacuated loops seen edge-on;
however after studying {\it Solar Dynamics Observatory} (SDO) Atmospheric Imaging
Assembly (AIA) 193, 131, and 94~\AA\ images, \citet{Savage12a} concluded that they were wakes
of retracting loops. Correlation tracking of the surrounding flows has shown that the surrounding
plasma is highly turbulent \citep{McKenzie13}. This is consistent with the
SUMER observations of high line-of-sight velocities in the supra-arcade fan along the edge of
SADs \citep{Innes03b,Doschek14}.

A  model, based on 3-D bursty reconnection,
demonstrated that teardrop-shaped downflows with trailing low density tails,
similar to SADs could be expected
from retracting loops \citep{Linton06,Guidoni11}. When the loops descend
through the supra-arcade fan, they
may drive shocks ahead and rarefactions behind \citep{Scott13} travelling with speeds
of the order of the local sound speed (350~\kms\ for 10~MK plasma).
The SAD model developed by \citet{Costa09}, \citet{Maglione11}, and
\citet{Cecere12}
assumes multiple reconnection sites in which the SADs are a consequence of
shocks and rarefactions bouncing back and forth in magnetic structures.

Recently, \citet{Cassak13} have pointed out that retracting loops or their
wakes would be filled too quickly by the surrounding plasma to explain their
relatively long lifetime (up to 20~min), and propose that the SADs are low density jets
produced by multiple reconnection sites along the current sheet. In their scenario,
the continual jet flow fills the SAD channel and prevents it from closing;
however their model places the voids below, not above, the flare
arcade.

An alternative jet-flow model, introduced by \citet{Asai04},
is that, rather than being a consequence of many reconnection jets, the SADs
 are the result of
instabilities at the head of a main jet \citep{Tandokoro05}.
The head of a low density jet pushing
into high density plasma above the arcade would be
Rayleigh-Taylor unstable. There are clear morphological similarities
between SADs and the Rayleigh-Taylor plumes and fingers recently identified in
high resolution solar
observations of quiescent prominences
\citep{Berger10, Berger11, Hillier12} and ejected filament plasma falling
back through the  corona \citep{Innes12}.

Distinguishing between the various SAD models is a challenge for the
observations. In this paper, we look carefully at  the early appearance
and growth of individual SADs at the top of the arcade fan because this is
where models differ.
For example, as described in an accompanying paper \citep{Guo14},
if SADs are caused by  instabilities at the interface between the
jet and fan plasma, they will exhibit structural evolution
at their heads whereas the descending loop
\citep{Savage12a, Scott13} and jet \citep{Cecere12, Cassak13} models predict
relatively stable teardrop-shaped heads. Also growing fan spikes
are a natural consequence of the instability model and are themselves expected to be
unstable, leading to the appearance of SADs at the top of growing fan spikes.
In addition, models differ in the
 relationship between  SADs and SADLs.
In the wake model SADLs are expected  near the front of all SADs,
whereas the Rayleigh-Taylor instability model of \citet{Guo14} predicts that SADs
are low density, reconnected jet plasma that can exist without a SADL at its
head.
However, as explained in the discussion section, the SADs are threaded by
flux rooted in opposite polarities in the photosphere so loops may appear along with
SADs if, for example, non-uniformities cause the SAD flux to break up.
SADLs have been observed in front of
some of the larger SADs as they approached the arcade \citep{Savage12a} but the
viewing angle for picking up SADs was not optimal and there was
considerable confusion along the
line-of-sight.
We test the SADL-SAD connection by taking advantage of an event that occurred
on the disk where we can see both SADs and SADLs with less
line-of-sight confusion and with a better orientation
for seeing faint SADLs.

\section{Observations}

\begin{figure*}
\includegraphics[width=\linewidth]{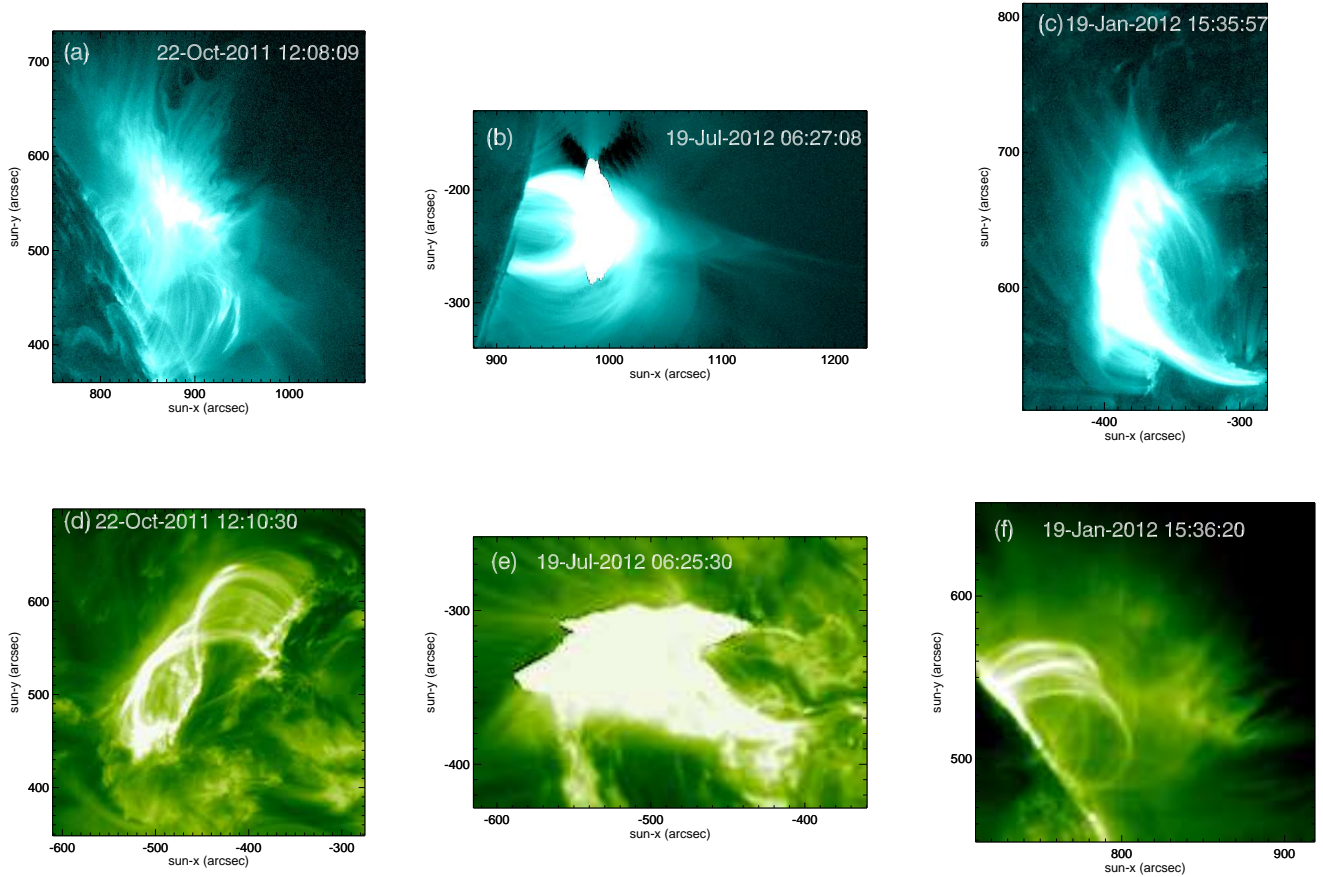}
\caption{(a)-(c) SDO/AIA 131~\AA\ channel images of the flares and
supra-arcades fans.
(d)-(f) Corresponding STEREO/EUVI 195~\AA\ filter images of the same flares from
(d) STEREO-A at 105\degree, (e) STEREO-A at 120\degree, (f) STEREO-B at
-113\degree\
Earth ecliptic longitude.}
\label{3flares}
\end{figure*}

We investigate SDO/AIA observations of SADs in three eruptive flares
with different arcade axis and therefore current-sheet orientations. The flares are
shown in Figure~\ref{3flares}. The top panels show the SDO/AIA images and
 the panels below, the flares seen from the STEREO
Extreme Ultraviolet Imager (EUVI)
 \citep{Howard08}.
STEREO consists of two identical spacecraft, STEREO-A (ahead) and STEREO-B (behind), in
heliocentric orbit at $\sim1$~AU that are increasing their separation from Earth
by $\sim22$\degree/year.
At the time
the STEREO spacecraft were between 100\degree\ and 120\degree\ from the
Earth-Sun line so SDO limb/disk flares appear as disk/limb flares
from STEREO.
The arcade axis after the M1.3  flare on 22 October 2011 was perpendicular to the line-of-sight
and so the current sheet was seen face-on  from SDO \citep{Savage12a}. The 2012 July 19 M7.7
 flare produced a bright, cusp-shaped arcade on the south-west limb.
As can be seen from the STEREO image, Figure~\ref{3flares}(e),
the viewing perspective from SDO was along the arcade axis
\citep{LiuW13,Liu13}. The M3.2  flare on 2012 January 19 occurred
close to disk center, at N28E13, and the arcade which ran north-south was
essentially seen from above from SDO.

AIA obtains full Sun images out to $\sim$1.3 R$_\odot$ with a spatial
resolution of 0.6\arcsec\ pixel$^{-1}$ and a cadence of 12~s in 10
EUV and UV channels \citep{Lemen12}.
SADs are visible in
all the hot AIA channel images (94, 131, 193, 335~\AA) but are best seen in the
131~\AA\ images. In these images the arcade fan is brightest and most extended.
The main contributor to the 131~\AA\ flare emission is \FeXXI\ which has a
formation temperature around 10-12~MK.

\subsection{2011 October 22}
The 131~\AA\ channel images show a large, rapidly evolving supra-arcade fan
 with a rich variety of SAD structures (see movie\_11oct22). Analysis
of images from this event led \citet{Savage12a} to conclude that the SADs were
not edge-on loops but rather the wakes behind rapidly retracting loops.
This is the event where they identified loop-like structures
 at the heads of some of the SADs as they approached the top of the
 underlying arcade.

\begin{figure*}
\includegraphics[width=0.9\linewidth]{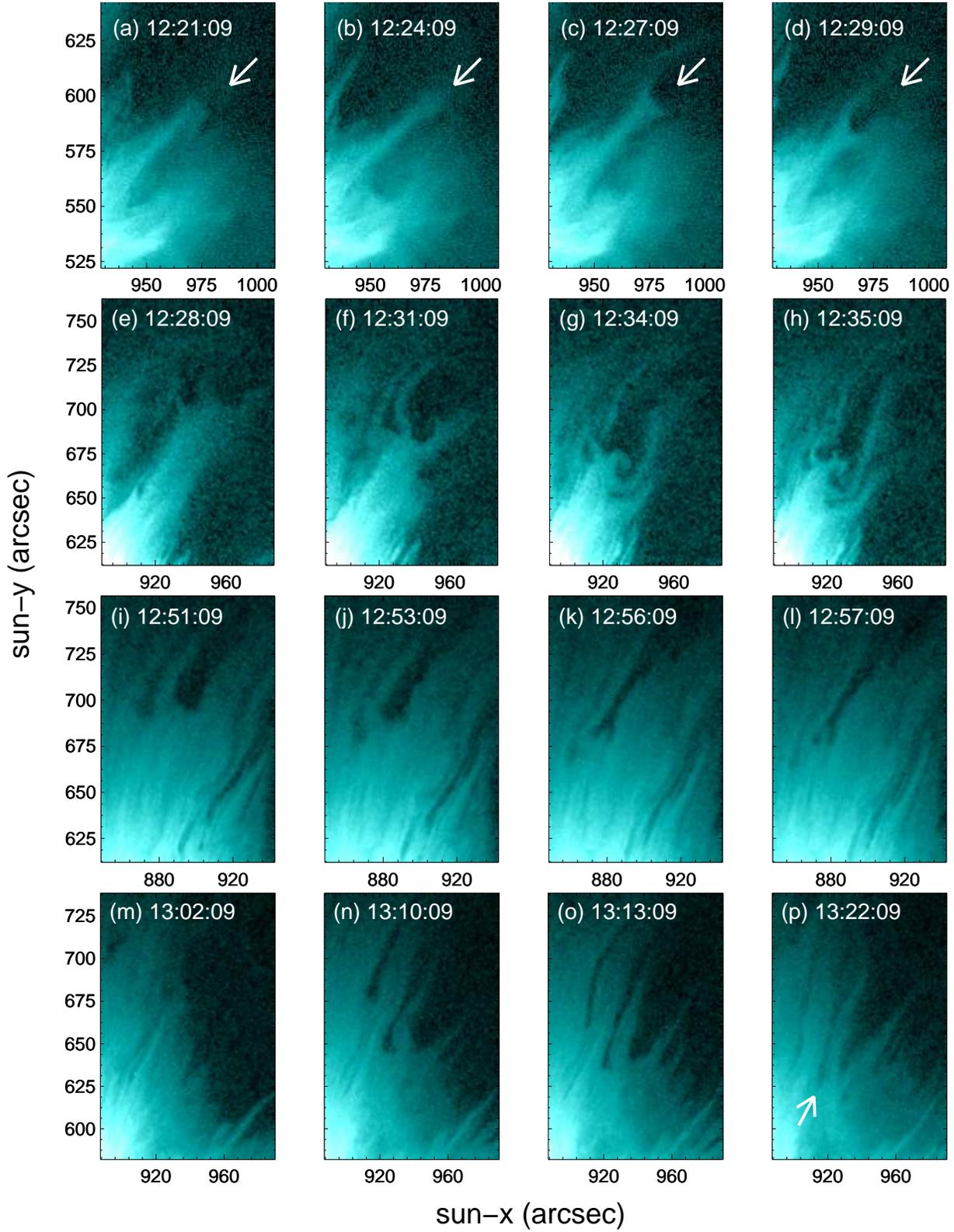}
\caption{SDO/AIA 131~\AA\ channel images showing the
 evolution of SADs in the supra-arcade fan of 2011 October 22.
(a)-(d) Appearance of a SAD at the top of a growing fan spike.
(e)-(h) The head of this large SAD develops prongs at the side of the head.
(i)-(l) The head of a smaller SAD also splits in two.
 (m)-(p) Long-lasting SADs that fade above the arcade with no indication of
brightening at their heads.
The white arrows in the top row point to a growing fan spike
where a SAD appeared at its apex. The white arrow in (p) points to the head of a fading
SAD.}
\label{sads_oct22}
\end{figure*}

\begin{figure*}
\includegraphics[width=0.9\linewidth]{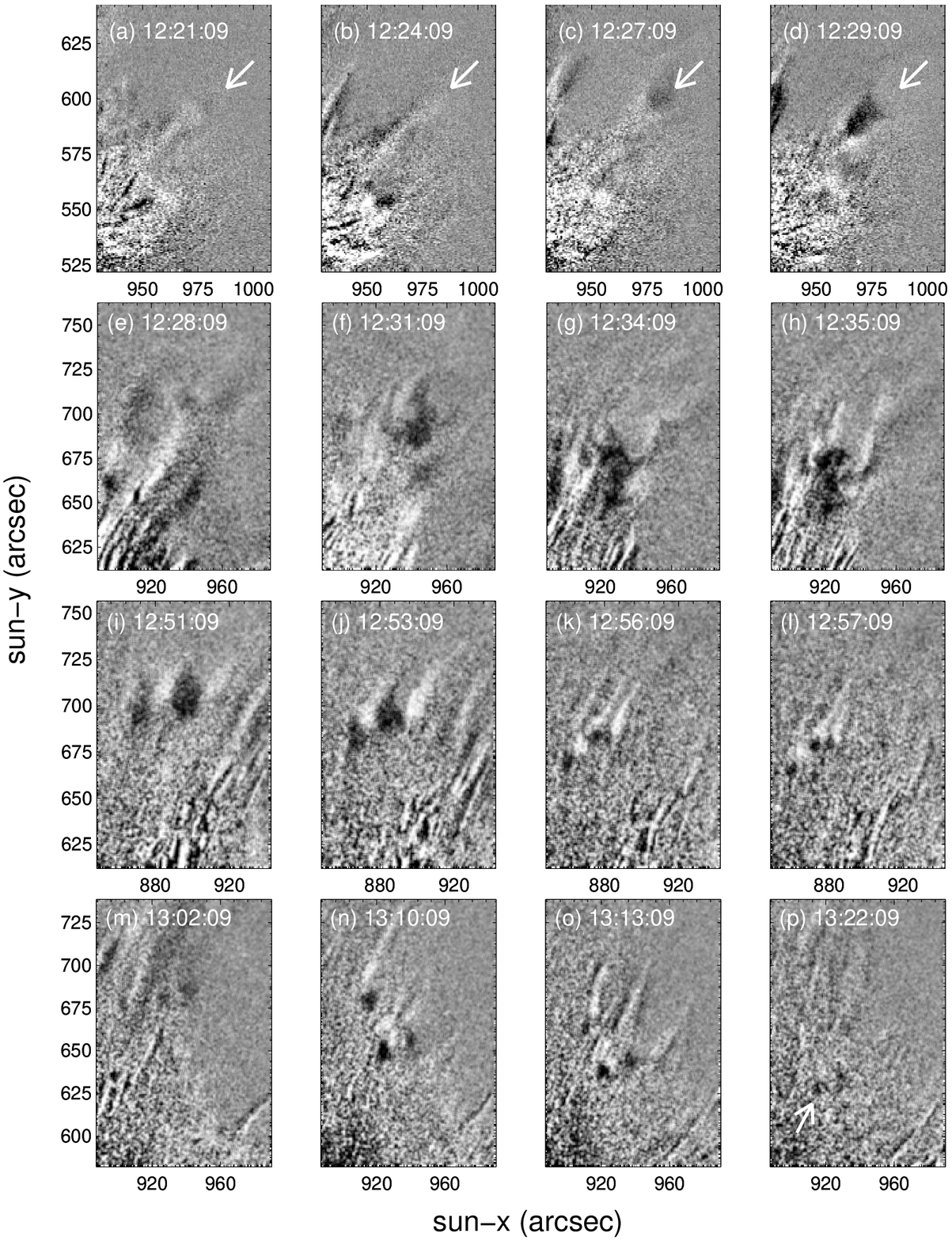}
\caption{SDO/AIA 131~\AA\ difference images at the same time as the images in
Figure~\ref{sads_oct22}. In each case, the subtracted image was taken 2 min earlier.
}
\label{sads_oct22_rd}
\end{figure*}

Here we concentrate on the
first appearance
and evolution of some of the clearest examples of SAD evolution.
Figure~\ref{sads_oct22} shows four time frames for four
groups of SADs. The corresponding difference images are
shown in Figure~\ref{sads_oct22_rd}.
 The first  SAD sequence
has been picked out because it shows a SAD appearing at the head of a growing
fan spike.
The growth of the spike, with a velocity 40~\kms, is clearly visible
between frames (a) and (b) in both the
intensity and difference images. The SAD, seen first in frame (c), appears at
the apex, causing the spike top to split.
Both the growing fan spike and the SAD appearance at the top are
predicted by the
simulations of SADs caused by instabilities in the exhaust
of reconnection jets \citep{Guo14}. An example is shown in the discussion
section,
Figure~\ref{sadsketch}(b)-(d).
The reconnection downflow occurs across the length of the arcade, and the whole
interface is unstable. The spikes represent the fastest growing modes.
The regions between the spikes are also filled with downflows
but these are not as visible because there is
less contrast with the fan emission.

The second row shows evolution at the
head of one of the wider SADs. The original teardrop-shaped head (f)
developed prongs (g) that grow (h) and later
separated from the main downflow,
creating 3 or 4 tadpole-like SADs (see movie\_11oct22).
The next example shows prong development at the head of a narrower SAD. In
this example, between images (k) and (l),
the dark `tail' remains stationary while the prongs detach and move
slowly downward.
In this supra-arcade fan, there are several examples of SADs
with evolving head structure. We presume it is common,
although it has not been reported elsewhere. Further investigation of
other well-observed events is required.

The fourth example shows two narrow SADs that remain visible
for at least 12 min although their widths are at most 1.5~Mm.
\citet{Cassak13} have suggested that if the SADs were wakes, they will  rapidly fill
with ambient plasma; however,  the ambient magnetic field
may compensate for the
lower plasma pressure in the wake preventing their disappearance.
 The long lifetime can also be explained
if SADs result from quasi-saturated
instabilities along
an interface between a hot, low-density jet,
 and denser, cooler arcade plasma because
the reconnection flow will
keep the SAD from filling with ambient plasma.
A more extreme example of a stationary SAD was reported by \citet{Hanneman14}.
The one they show is also fork-shaped, and it remained at the same position for about one hour.
We also note that the last two narrow SADs simply slowed down and faded in the
bright fan plasma. There was no indication of either a loop or enhanced emission
 at their heads, as has been predicted by some retracting loop models
 \citep{Scott13}.


\subsection{2012 July 19}
\begin{figure*}
\includegraphics[width=0.9\linewidth]{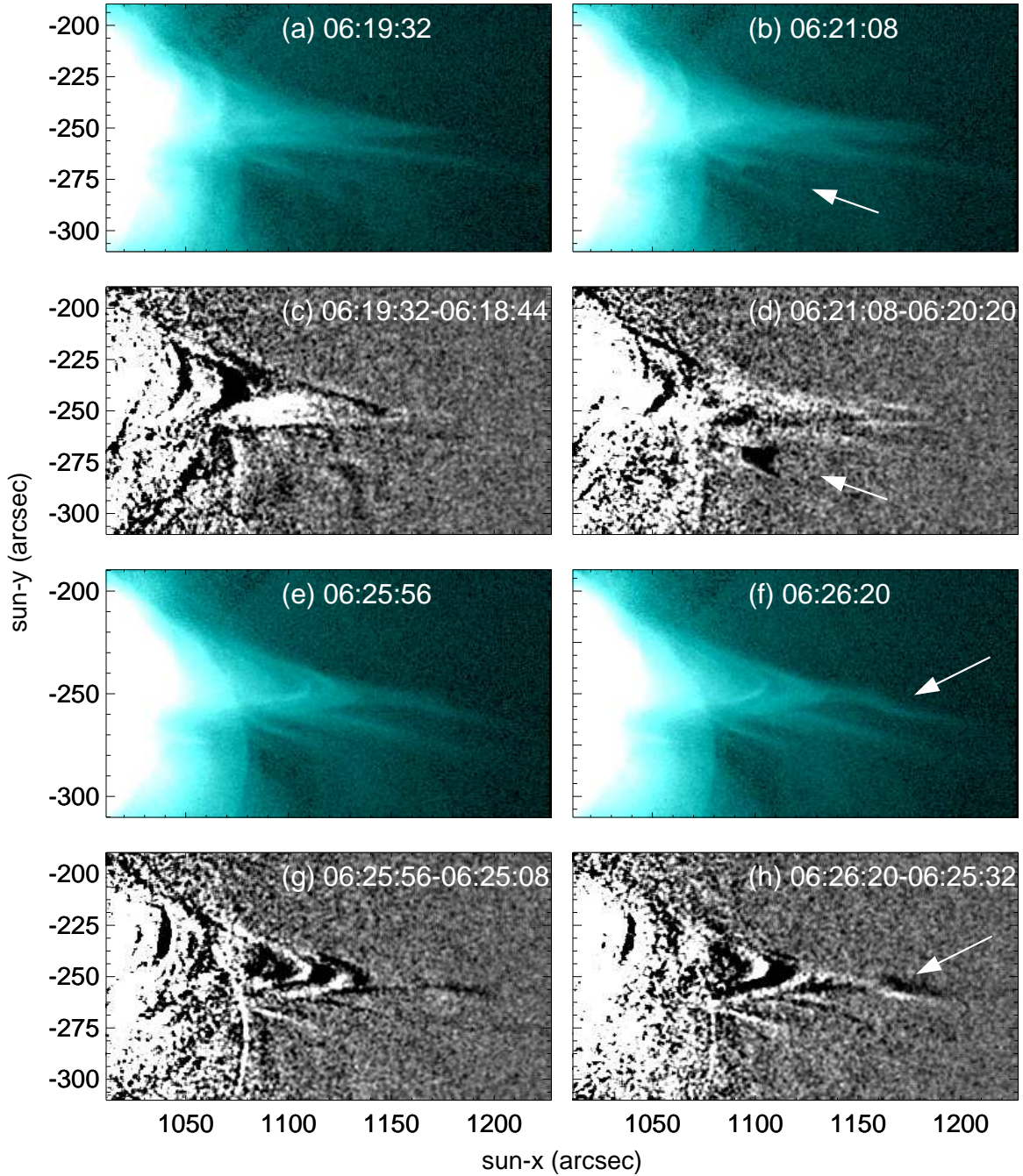}
\caption{SDO/AIA 131~\AA\ intensity and  difference images showing the
 evolution of SADs in the supra-arcade fan of 2012 July 19.
(a)-(b) Formation of a SAD (white arrow) at the top of a fan spike
above a cusp arcade.
(c)-(d) Corresponding difference images.
(e)-(f) SAD formation below the main current sheet (white arrow).
(g)-(h) Corresponding difference images.}
\label{sads_jul19}
\end{figure*}

This bright, cusp-shaped event has been analysed by \citet{LiuW13} and
\citet{Liu13} who have described many aspects related to the dynamics of the
bright fan emission, including coronal hard X-ray and microwave sources, fast
retracting loops, and SADs. As can be seen in the online movie, movie\_12jul19,
many SADLs were seen during the initial phase of arcade formation.  The earliest SADs,
appeared later than the first SADLs. They were analysed by \citet{Liu13}, and
are shown again in Figure~\ref{sads_jul19}.
The first SAD, indicated by a white arrow in Figure~\ref{sads_jul19}(b) and (d),
appeared at the top of one
of the fan spikes. This was a very narrow spike
and the SAD appeared almost exactly at its apex, suggesting that the presence
of the spike may be an important aspect of the SAD's formation.
This however requires a detailed study of the statistics of SADs and
spikes. It is also possible that the SAD forms higher in the corona and
falls onto the fan spike.

In the second
example, Figure~\ref{sads_jul19}(f) and (h), the main fan spike moved
sideways, pushed by downflowing plasma. \citet{Liu13} has shown that this
downflow appeared to be related to a slowly retracting
SADL underneath.
 There is however considerable line-of-sight confusion because emission is integrated
along the whole length of the arcade, so it is not clear how the SAD and SADL are
related.
The distance between the SAD and SADL is rather large for the SAD to be the wake
of the SADL, so \citet{Liu13} suggested that the SAD is a twisted flux rope contracting
onto the arcade.

\subsection{2012 January 19}
\begin{figure*}
\includegraphics[width=0.9\linewidth]{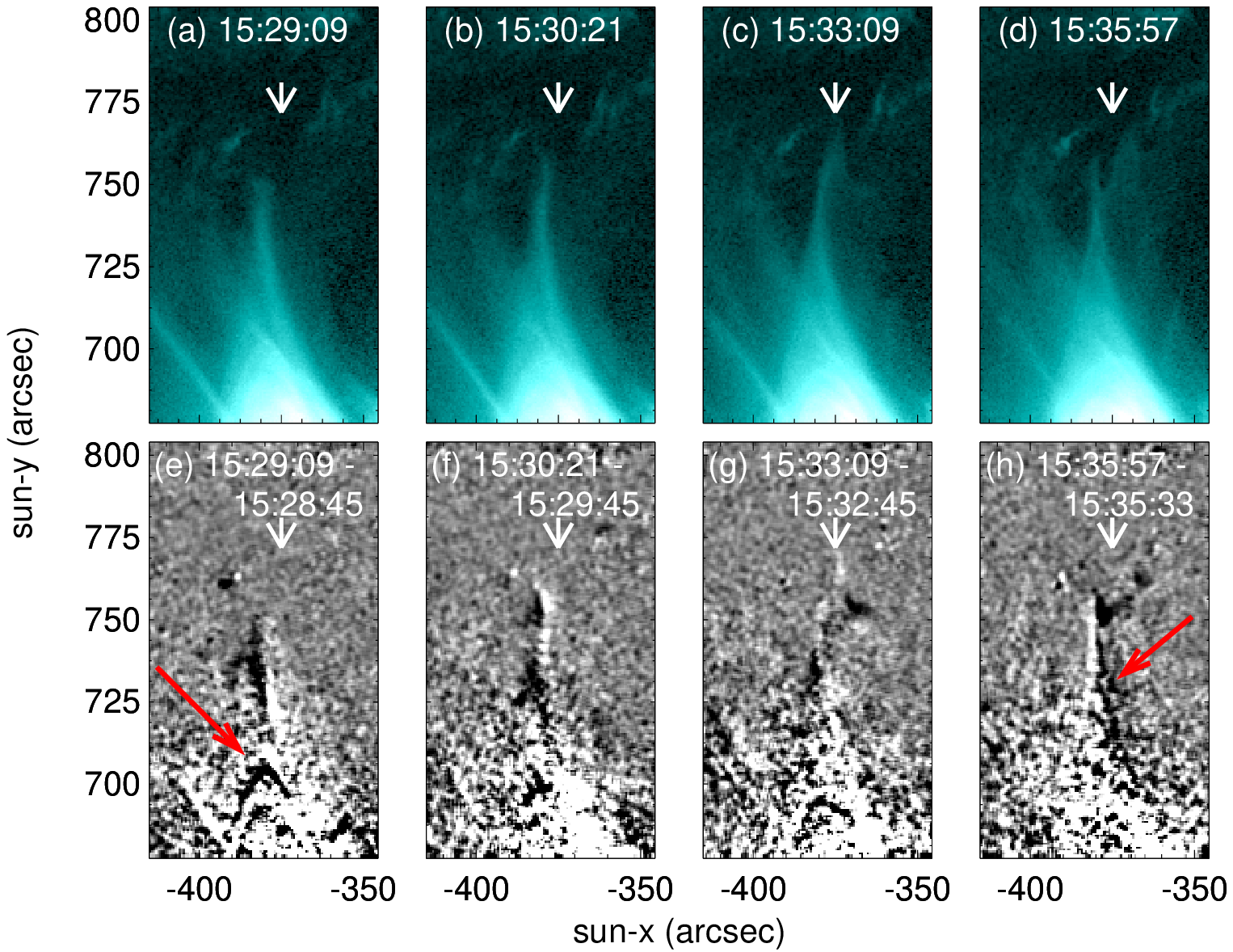}
\caption{SDO/AIA 131~\AA\ intensity and difference images showing the
 evolution of SADs in the supra-arcade fan of 2012 January 19.
(a)-(d) Formation of a SAD (white arrow) at the top of a fan spike
above a cusp arcade.
(e)-(h) Corresponding difference images. The white arrows indicate
the fan spike where the SAD formed. The red arrows indicate SADLs.
}
\label{sads_jan19}
\end{figure*}

This event gives a less confused view of SADs forming
below an edge-on current sheet
 because the supra-arcade fan is projected against the disk,
and viewed at a slight angle so that the northern
part has the dark disk as background while
the rest is superimposed on the bright flare arcade.
The STEREO images of this event show that there were many SADs along the whole arcade
({\it e.g.} Figure~\ref{3flares}(f)).
As shown in the online movie, movie\_12jan19,
individual SADs can clearly be
picked out in AIA images at the northern end of the arcade.
Again several SADs are seen at the top of fan spikes.
One example is shown in Figure~\ref{sads_jan19}.
Initially the fan spike, indicated with a white arrow, grows upward with velocity
about 50~\kms. The growth of
the spike  is most visible in the difference
images. In Figure~\ref{sads_jan19}(g) the top of the spike is white,
indicating increased emission at the top of the spike just before the SAD appears
(Figure~\ref{sads_jan19}(h)).

\begin{figure*}
\includegraphics[width=0.9\linewidth]{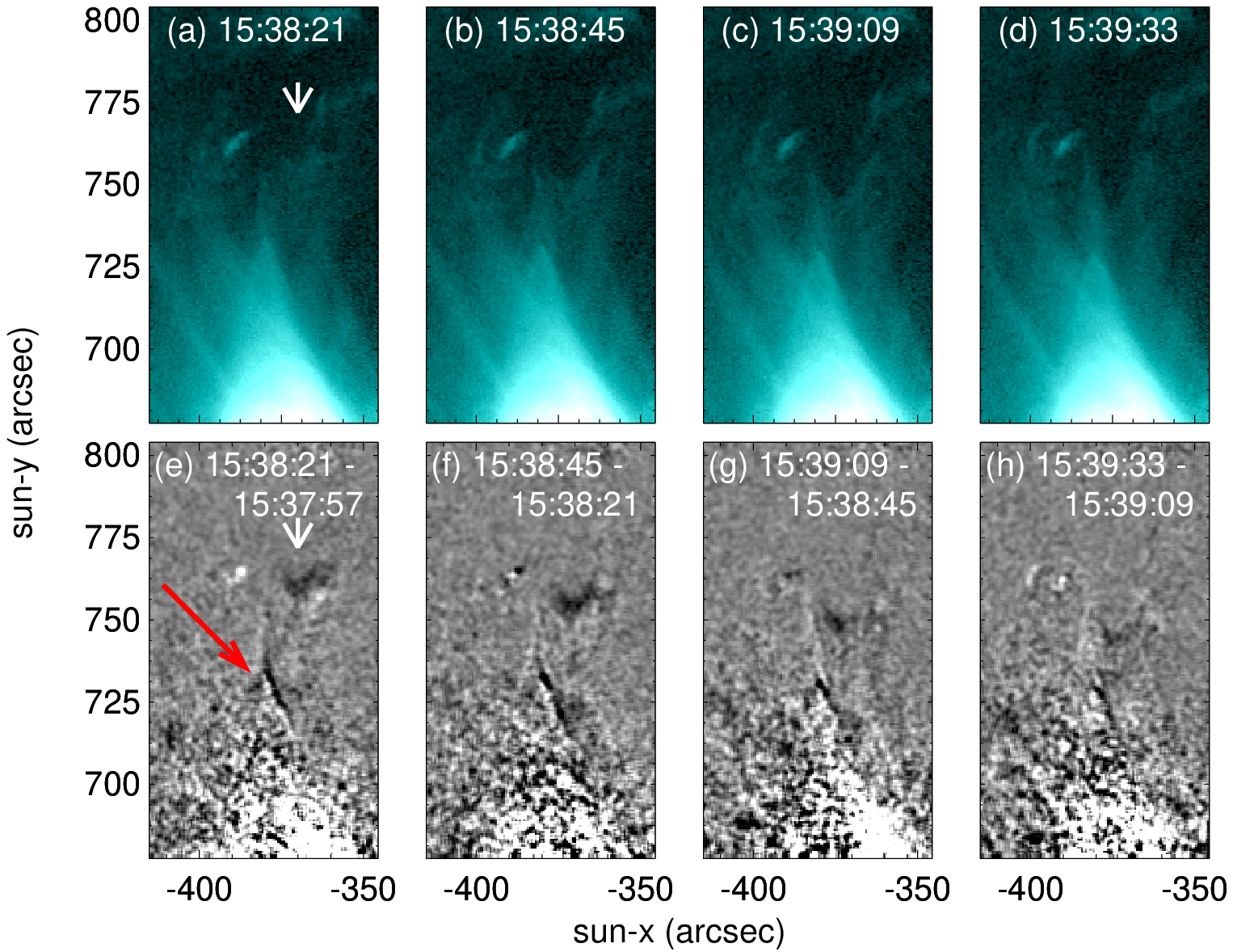}
\caption{SDO/AIA 131~\AA\ intensity and difference images showing the
 evolution of SADs in the supra-arcade fan of 2012 January 19.
(a)-(d) SAD (white arrow) with no SADL at its head.
(e)-(h) Corresponding difference images.
The red arrow indicates the nearest SADLs.
}
\label{sads_jan19_2}
\end{figure*}

The advantage of this event over the previous two discussed, is that the
 relationship between SADLs and SADs can be seen much more clearly.
SADLs can be unambiguously identified since they appear broad-side rather than edge-on,
 and there is less line-of-sight confusion
because one is only looking through a small part of the arcade not the whole thing.
 In Figures~\ref{sads_jan19} and \ref{sads_jan19_2},
SADLs are indicated with red arrows. Figure~\ref{sads_jan19}, shows a SADL
that appeared at about the same time and about 25\arcsec\ below a SAD. There may be a connection
between the two because the difference image suggests that the SADL appears at the base of the
spike when the SAD appeared at its top. This is similar to the association
between  the SAD and SADL
 recorded by \citet{Liu13} in the 2012 July 19 event.
 In the second example,
 shown in Figure~\ref{sads_jan19_2},
 the difference images
show no connection between the two, and  the SAD is descending faster than the SADL.
There is also no bright 131~\AA\ emission at the head of the SAD.
Several other examples of SAD and SADL evolution
can be seen in the movie, movie\_12jan19rd. There are both connected and
separate SADs and SADLs.
In the movie, SADs generally appear above the height where SADLs are seen.

\section{Discussion}
By looking at the appearance  of SADs in the supra-arcade fans of three eruptive
flares with different orientations of their arcade axes,
we find that:  (i) the heads of SADs evolve, sometimes splitting into narrower
SADs;
(ii) a retracting loop, SADL, is sometimes seen below a SAD but there is no strong evidence in our datasets that
SADs and SADLs are directly  connected;
(iii) several SADs appear at the top of rising spikes;
(iv) SADs do not necessarily have bright emission at their heads.
We also note two other features that have been much
discussed and poorly understood: (i) the relatively long lifetime
of narrow SADs and (ii) their slow downward velocity, typically 25-50~\kms,
as measured from EUV images.

\begin{figure*}
\includegraphics[width=0.95\linewidth]{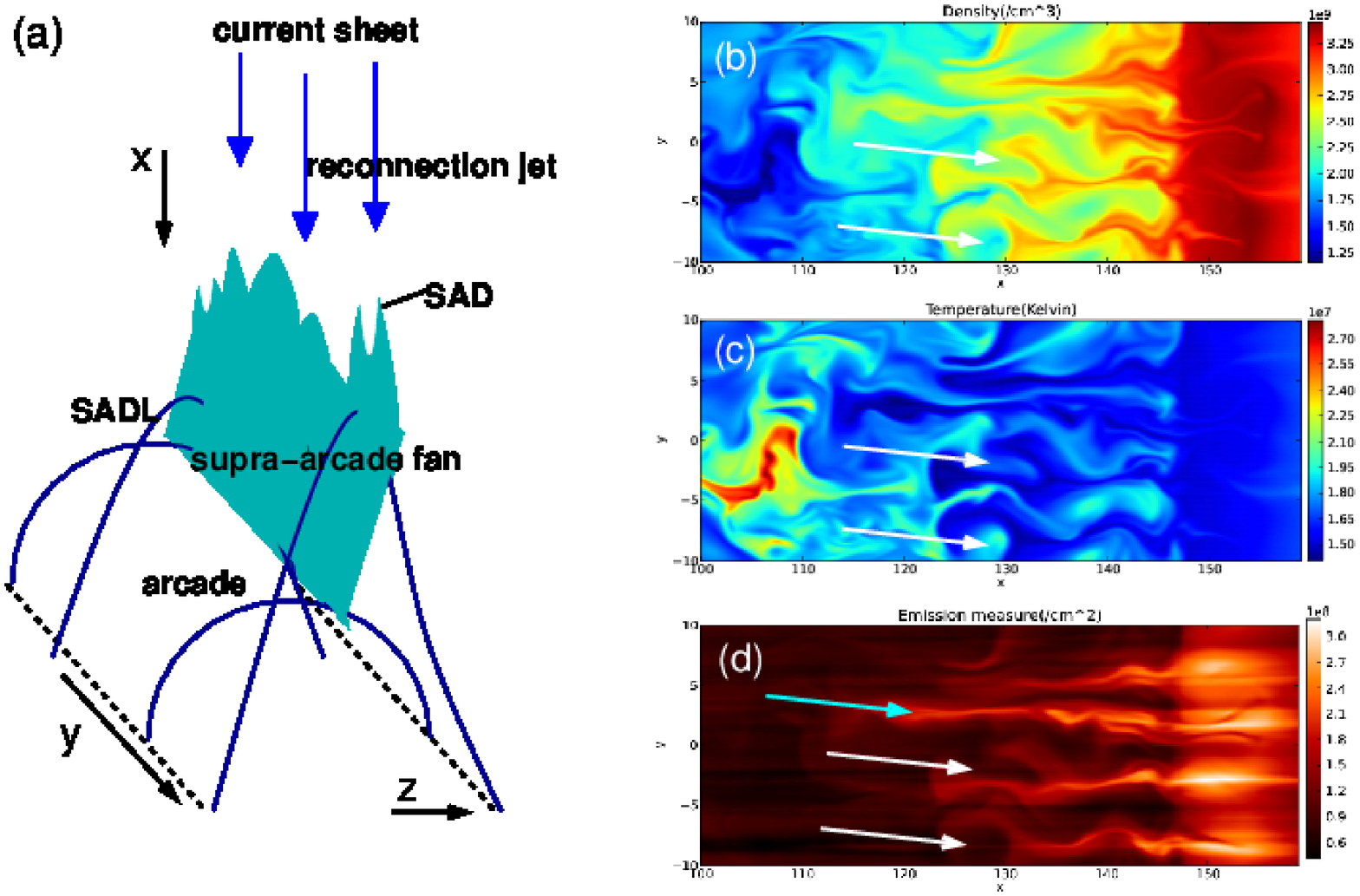}
\caption{(a) Sketch of a supra-arcade showing the configuration of the reconnection jet
 and supra-arcade
fan. (b)-(d) Simulations of instabilities at the head of a reconnection jet: (b) density;
(c) temperature; (d) emissivity in the AIA 131 \AA\ channel. Temperature and density are
shown along the plane passing through the center of the current sheet(Z=0),
and emissivity
is integrated across the Z-direction. The directions X, Y, Z are indicated in (a).
The white arrows in (b)-(d)
point to downflows forming on top of rising high-density fingers.  The turquoise arrow in (d) points to a fan spike.}
\label{sadsketch}
\end{figure*}

As discussed in \citet{Guo14}, these phenomena
 can be naturally explained in the context of instabilities in the exhaust
of a reconnection jet. Figure~\ref{sadsketch}(a) illustrates the configuration.
 Following the CSHKP model of
eruptive flares \citep{Car64,Sturrock66,Hira74,KP76}, a current
sheet forms behind a coronal mass ejection (CME). The core of
the CME is usually an elongated filament, so the eruption generates a long,
irregular current sheet behind the erupting filament.  Reconnection along the
current sheet leads to outflows and the formation of an arcade
 into which plasma continues to accumulate leading to
 a high density region below the
outflowing jet.
The outflows occur along the length of the current sheet and
create a broad jet of low density, high-speed plasma.
When this plasma encounters the higher density plasma that has filled the
region between the jet head and the arcade, the interface will be unstable to modes
of the Rayleigh-Taylor type
(including interchange/ballooning modes, if the plasma conditions allow).
A snapshot of the density, temperature and corresponding 131\AA\ intensity
of plasma along the interface, from one of the simulation runs described in \citet{Guo14}
is shown in Figure~\ref{sadsketch}(b)-(d).
As can be seen the instability
results in
fingers of high
temperature and low density penetrating into the underlying higher density plasma.
The low density fingers, which are part of the reconnection outflow,
are threaded by reconnected flux with footpoints rooted in
opposite polarities on the solar surface.
The difference between
the interpretation of SADs presented here and the loop picture
\citep{Linton06,Savage12a,Scott13} is that here the SADs are due to an
instability in the reconnection outflow whereas in the loop model the SADs are
a consequence of retracting loops.
It may be that as more of the reconnected flux piles up
towards the top of the arcade, chromospheric
evaporation causes the head of SADs to fill with plasma,
and thus create the SADLs. To simulate this would require more realistic
boundary conditions and additional physics.
These snapshots
were chosen because they show that dark tadpole-shaped structures (SADs) appear
at the apex of the high density plumes (spikes),
as seen in the observations.
The most obvious discrepancy with observation is that the
ratio between the finger separation to their length is smaller than
typically seen in SADs.
In practice, the separation is probably governed by inhomogeneities in the
arcade fan, the strength of the magnetic field in the Y-direction, and the
reconnection process
which may be subject to two-fluid or kinetic effects outside
of the scope of the MHD model.
Another strong point of the model is that it explains
the low plane-of-sky velocity of SADs seen in EUV images,
 and their relatively long lifetime.

Other explanations for SADs, in particular the retracting loop model, may be
configured to reproduce the splitting at the heads of the SADs if one considers
multiple loops, and the lack of 131~\AA\ brightening at their head by arguing
that it is outside the bandpass of the observations, or that poor contrast
at the top of the fan hides the SADL/shock emission.
We hope that the aspects of SAD evolution highlighted here will motivate
further modelling, especially of the retracting loop models to test these
hypotheses.

The instability model explains the principle features of
SADs discussed here, and provides a coherent picture of reconnection outflows,
the formation of arcades, supra-arcade fans, and SADs.
We  conclude therefore
that SADs appear due to instability at the
interface where a fast, low density, reconnection jet
encounters higher density supra-arcade fan plasma.

\vskip 1.0cm

We thank the referees for constructive comments. This work is supported by the Department of Energy,
Grant No. DE-FG02-14-07ER46372, under the auspices of the Center for
Integrated Computation and Analysis of Reconnection and Turbulence (CICART),
the National Science Foundation, Grant No. PHY-0215581
(PFC: Center for Magnetic Self-Organization in Laboratory and
Astrophysical Plasmas), NASA Grant Nos. NNX09AJ86G and
NNX10AC04G, and NSF Grant Nos. ATM-0802727, ATM-090315 and AGS-0962698.


\end{document}